\documentclass[a4paper,11pt]{article}
\pdfoutput=1 
\usepackage{jheppub} 
\usepackage[T1]{fontenc}
\usepackage[italian,english]{babel}
\usepackage{hyperref}
\hypersetup{
	colorlinks=true,
	linkcolor=[rgb]{0.0,0.42,0.89},
	filecolor=magenta,      
	urlcolor=[rgb]{0.0,0.42,0.89},
	citecolor=[rgb]{0.64,0.0,0.0},
}
\usepackage{ifpdf}
\usepackage{subfigure}
\usepackage{gensymb}
\usepackage{amssymb}
\usepackage{amsfonts}
\usepackage{slashed}
\usepackage{epsf}
\usepackage{rotating}
\usepackage{graphicx}
\usepackage{amsmath}
\usepackage{fancyhdr}
\usepackage{lineno}
\usepackage{babel, blindtext}
\usepackage{graphics}
\usepackage{color}
\usepackage{physics}
\usepackage{multirow}
\usepackage{pstricks}
\usepackage{xcolor}
\usepackage{multirow}
\usepackage{hhline}
\usepackage{cancel}
\usepackage{framed}
\usepackage{mathtools}
\usepackage{diagbox}
\usepackage{float}
\usepackage{orcidlink}
\usepackage{ulem} 
\usepackage[lmargin=1.7in,bmargin=0.2in,footskip=0.5in,total={6.9in,9.5in}]{geometry}

\newcommand{\lsim}{\mathrel{\mathop{\kern 0pt \rlap
			{\raise.2ex\hbox{$<$}}}
		\lower.9ex\hbox{\kern-.190em $\sim$}}}
\newcommand{\gsim}{\mathrel{\mathop{\kern 0pt \rlap
			{\raise.2ex\hbox{$>$}}}
		\lower.9ex\hbox{\kern-.190em $\sim$}}}

\newcommand{\be}{\begin{equation}}
	\newcommand{\ee}{\end{equation}}
\newcommand{\bea}{\begin{eqnarray}}
	\newcommand{\eea}{\end{eqnarray}}

\title{\boldmath Entanglement Signatures of Kinetic-Mixing Portals in Dark Monopole Scattering}

\author[a]{Shilpa Jangid \orcidlink{0000-0001-6307-1234},}
\author[b]{Hiroshi Okada \orcidlink{}}

\affiliation[a]{Shiv Nadar IoE Deemed to be University, Gautam Buddha Nagar, Uttar Pradesh, 201314, India}
\affiliation[b]{Department of Physics, Henan Normal University, Xinxiang 453007, China}

\emailAdd{ shilpajangid123@gmail.com, hiroshi3okada@htu.edu.cn}

\preprint{}

\abstract{ We examine the interaction between the Standard Model (SM) fermions and the topological dark magnetic monopoles mediated via a kinetic-mixing portal to investigate the generation of quantum entanglement in particle-portal scattering. While the conventional phenomenology only takes into account total cross-sections, decay widths, and missing-energy signatures, we employ an information-theoretic approach to explain the scattering event. We quantify the quantum correlations transported over the portal boundary by calculating the Von Neumann entropy ($S_{\text{ent}}$), and the subsystem purity deterioration ($\gamma < 1$), analytically. Our results demonstrate that the non-perturbative core form factor of the dark monopole controls its high-energy momentum transfer, and that the Von Neumann entropy increases quadratically with the topological magnetic charge ($g_m^2$) and the kinetic-mixing parameter ($\epsilon^2$). This paradigm provides an information-theoretic "microscope" to restrict portal parameter spaces and explore underlying topological structures without using traditional energy signatures by establishing quantum decoherence and purity loss as new, complementary observables.}


\begin{document}

\maketitle
\flushbottom
\section{Introduction}
The search for physics beyond the Standard Model (SM) has focused more and more on weakly interacting hidden or dark sectors \cite{Essig:2013lka, Goodsell:2009xc}. The kinetic-mixing portal parameterized by a small dimensionless coupling $\epsilon$ between the standard hypercharge gauge boson and a hidden $U(1)_D$ dark photon $A'$ is still a fundamental benchmark in particle phenomenology among the many portals proposed to mediate interactions between visible matter and hidden fields \cite{Holdom:1985ag}. Precision electroweak measurements, beam dump experiments, collider missing-energy channels, and astrophysical cooling constraints have all been used in extensive experimental searches and theoretical analyzes of kinetic mixing \cite{Essig:2013lka}. These phenomenological studies have historically only used classical or semi-classical observable metrics, like missing-momentum distributions, decay widths, total scattering cross-sections, and invariant mass resonance peaks. However, the multi-scale dynamics and internal zero-modes introduce complex quantum correlations when scattering processes involve non-perturbative topological configurations, such as dark magnetic monopoles resulting from spontaneous symmetry breaking in the hidden sector \cite{tHooft:1974kcl,Polyakov:1974ek,Preskill:1984gd}, which can be averaged over or washed out when viewed exclusively through inclusive probability rates. In this work, we present an information-theoretic framework to study the quantum entanglement between dark sector topological defects and visible Standard Model (SM) particles in order to examine portal-mediated scattering. In addition to transferring four-momentum, the interaction process of a Standard Model (SM) fermion scattering off a dark magnetic monopole through a kinetic-mixing portal dynamically entangles the visible particle's internal spin and kinematic degrees of freedom with the unobserved internal zero-modes and quantum recoil states of the monopole core. Using the partial trace formalism \cite{KRAUS1971311}, we treat the scattered visible fermion as an open quantum system and integrate out the inaccessible hidden-sector degrees of freedom to obtain a reduced density matrix $\rho_{\text{vis}}$. We calculate the Von Neumann entropy ({$S_{ent}$})
\cite{Vidal:2002zz,Calabrese:2004eu} to quantify the ensuing quantum information leakage and establish a direct analytical connection between quantum correlations and fundamental portal characteristics \cite{Holdom:1990xp}. We show that the non-perturbative core form factor of the dark monopole controls momentum transfers and that entanglement deterioration and purity loss scale quadratically with both the kinetic-mixing parameter ($\epsilon^2$) and the topological magnetic charge ($g_m^2$). This method offers a supplementary, coherence-sensitive diagnostic tool that may be used independently of conventional energetic signs to confine portal parameter spaces and probe hidden topological structures. This is the structure of the rest of the paper. In order to diagonalize the kinetic mixing, Section~\ref{low-energy} describes the fundamental Lagrangian formulation and field redefinition \cite{Holdom:1985ag}. The portal-mediated scattering amplitude and the non-perturbative dark monopole vertex structure are derived in Section~\ref{scattering} \cite{tHooft:1974kcl,Polyakov:1974ek}. Subsection~\ref{Partial trace} assesses the subsystem purity deterioration \cite{KRAUS1971311} and builds the reduced density matrix using the partial trace operation. The  Von Neumann entropy is calculated analytically and its main physical ramifications and potential phenomenological traces are described in Section~\ref{entropy} \cite{Vidal:2002zz,Calabrese:2004eu}.
\label{sec:intro}   

\
\section{UV Completion and Topological Monopole Genesis}\label{low-energy}
To provide a solid theoretical basis for the dark magnetic monopole scattering framework, we propose that the dark sector originates from a non-abelian gauge group at high energies \cite{tHooft:1974kcl,Polyakov:1974ek}. We assume a $SU(2)_D$ gauge theory equipped with an adjoint dark Higgs field $\Phi^a$ ($a = 1, 2, 3$), regulated by the ultraviolet Lagrangian as follows:
\bea
\mathcal{L}_{\text{UV}} = -\frac{1}{4} F^a_{\mu\nu} F^{a\mu\nu} + \frac{1}{2} (D_\mu \Phi^a)(D^\mu \Phi^a) - V(\Phi).
\eea
In this case, $F^a_{\mu\nu} = \partial_\mu A^a_\nu - \partial_\nu A^a_\mu + g_D \epsilon^{abc} A^b_\mu A^c_\nu$ is the non-abelian field strength tensor, and $D_\mu \Phi^a = \partial_\mu \Phi^a + g_D \epsilon^{abc} A^b_\mu \Phi^c$ is the covariant derivative acting on the adjoint scalar field. The Mexican-hat scalar potential that regulates the kinetics of spontaneous symmetry breaking is defined as follows:
\bea
V(\Phi) = \frac{\lambda}{4} (\Phi^a \Phi^a - v_D^2)^2.
\eea
The standard Georgi-Glashow mechanism \cite{tHooft:1974kcl,Polyakov:1974ek} causes the original non-abelian symmetry to spontaneously break down to the residual abelian subgroup when the dark Higgs field acquires a non-zero vacuum expectation value along the third group generator, $\langle \Phi^a \rangle = v_D \delta^{a3}$.
\bea
SU(2)_D \to U(1)_D.
\eea
$SU(2)_D / U(1)_D \cong S^2$ is a non-trivial vacuum manifold isomorphic to the two-sphere that results from this symmetry-breaking pattern. Thus, the non-trivial mapping from the spatial sphere at infinity to this vacuum manifold, which is categorized by the second homotopy group \cite{Kibble:1976sj,Manton:2004tk}, strictly guaranties the topological stability of the resulting magnetic monopoles:
\bea
\pi_2(SU(2)_D / U(1)_D) \cong \pi_1(U(1)_D) = \mathbb{Z}.
\eea
This topological non-triviality naturally results in stable 't Hooft-Polyakov monopoles with a quantized topological magnetic charge \cite{tHooft:1974kcl,Polyakov:1974ek}:
\bea
g_m = \frac{4\pi n}{g_D} \quad (n \in \mathbb{Z}).
\eea
The mass spectrum generated after this spontaneous symmetry breaking consists of a dark Higgs radial mode $\Phi^3$ with mass $M_H = \sqrt{2\lambda} v_D$, a light $U(1)_D$ gage field $A^3_\mu \equiv A'_\mu$ that is massless at this primary stage, heavy non-abelian gauge bosons $A^{1,2}_\mu$ acquiring masses of $M_W = g_D v_D$, and a heavy topological monopole soliton whose mass scales as $M_{\text{mono}} \sim \frac{4\pi v_D}{g_D}$ \cite{Goddard:1977da}. These characteristic energy scales create a natural mass hierarchy such that $M_W, M_H \ll M_{\text{mono}}$ in the weak-coupling limit where $g_D \ll 1$, guaranteeing the validity of characterizing the topological defects as heavy, localized solitons.

\subsection{Low-Energy Effective Theory}
The heavy non-abelian gauge fields $A^{1,2}_\mu$ and scalar fields $\Phi^{1,2,3}$ are integrated out from the path integral \cite{Collins:1984xc} for scattering processes that take place at energies considerably below the heavy particle threshold where $E \ll M_W, M_H$. This results in the effective Lagrangian at low energy:
\bea
\mathcal{L}_{\text{EFT}} = -\frac{1}{4} F'_{\mu\nu} F'^{\mu\nu} + \mathcal{L}_{\text{monopole}} + \mathcal{O}\left(\frac{1}{M_W^2}\right),
\eea
where $\mathcal{L}_{\text{monopole}}$ defines the topological soliton and its non-perturbative core structure and $F'_{\mu\nu}$ represents the field strength tensor of the residual abelian $U(1)_D$ gage field $A'_\mu \equiv A^3_\mu$. In order to maintain the stability of the monopole core structure, a secondary mechanism, such as a second-stage dark Higgs or Stueckelberg mechanism, generates the physical mass of the dark photon $m_{A'}$, assuming the strict hierarchy $m_{A'} \ll M_W$ \cite{tHooft:1974kcl,Polyakov:1974ek}. Since the topological monopole background and the low-energy effective abelian gauge field ($A'_\mu$) are established in Section~\ref{low-energy}, the following section describes how this dark sector interacts with visible-sector fermions and gains its physical mass.

\section{The Model and Lagrangian:}\label{mixing}
We extend the Standard Model to include the residual abelian dark sector and its portal interactions with visible fermions, building on the low-energy effective theory developed in Section~\ref{low-energy}, where heavy non-abelian fields are integrated out below the threshold $E \ll M_W, M_H$. 
A potent probe for portals such as kinetic mixing can be found by examining quantum entanglement produced by particle scattering in hidden sectors \cite{Holdom:1985ag,Dienes:1996zr}. The kinetic-mixing portal mediates non-trivial quantum correlations across the sectors when magnetic monopoles charged under a dark $U(1)_D$ (or mixed $U(1)_{B-L}$) scatter off visible-sector fermions or standard magnetic/electric charges.\\

To formalize this arrangement, we supplement the Standard Model (SM) with an additional sequestered abelian gauge group $U(1)_D$ (or $U(1)_{B-L}$), which interacts solely through the kinetic-mixing portal.

\subsection{ Gauge Sector and Kinetic Mixing}
The Lagrangian density for the pure gauge and kinetic mixing is defined as:
\bea
\mathcal{L}_{\text{gauge}} = -\frac{1}{4} B_{\mu\nu} B^{\mu\nu} - \frac{1}{4} F_{\mu\nu}' F^{\prime\mu\nu} + \frac{\epsilon}{2} B_{\mu\nu} F^{\prime\mu\nu}, 
\eea 
where $B_{\mu\nu} = \partial_\mu B_\nu - \partial_\nu B_\mu$ and $F_{\mu\nu} = \partial_\mu A_\nu' - \partial_\nu A_\mu'$ are the field strength tensors of the visible hypercharge and dark abelian gauge fields, respectively. $\epsilon$ is the dimensionless kinetic-mixing parameter \cite{Holdom:1985ag,Dienes:1996zr}.

\subsection{ Dark Photon Mass Generation}
A secondary symmetry-breaking mechanism (like a Stueckelberg mechanism or a sequestered dark Higgs scalar $S$) produces the mass term \cite{Ruegg:2003ps,Pospelov:2007mp} in order to preserve the stability of the topological monopole core established in the UV completion while giving the dark photon a non-zero physical mass $m_{A'} \ll M_W$: 
\bea 
\mathcal{L}_{\text{mass}} = \frac{1}{2} m_{A'}^2 A'_\mu A^{\prime\mu} + \left( D_\mu S \right)^\dagger \left( D^\mu S \right) - V(S), 
\eea 
where the mass term (such as $m_{A'} = 2 g_D v_D$ for a dark charge $Q_D(S) = 2$) is generated by $\langle S \rangle = v_D / \sqrt{2}$, leaving behind a physical scalar radial mode that decouples in the heavy-mass limit.

\subsection{ Fermion Couplings and Topological Dark Monopoles}
According to \cite{Jaeckel:2010ni}, the extended covariant derivative for visible fermions $f$ that incorporates the dark gauge coupling $g_D$ and dark charge $q_D$ is as follows:
\bea 
D_\mu = \partial_\mu - i g_1 Y B_\mu - i g_2 T^a W^a_\mu - i g_3 G^A_\mu - i g_D q_D A'_\mu,
\eea
The kinetic-mixing term $\epsilon B_{\mu\nu} F^{\prime\mu\nu}$ induces effective off-diagonal dyonic charges upon mass-eigenstate diagonalization when the hidden sector admits topological defects, namely dark magnetic monopoles with topological magnetic charge \cite{Dirac:1931kp,Wu:1975es} $g_m = \frac{2\pi n}{g_D}$~\footnote{The normalization of the magnetic charge depends on the convention 
for the Dirac quantization condition. In Section \ref{low-energy}, we use the convention 
$g_m = 4\pi n/g_D$ following the 't Hooft-Polyakov monopole solution. In 
Section \ref{mixing}, we adopt $g_m = 2\pi n/g_D$ for consistency with the standard 
Dirac quantization $e g_m = 2\pi n$. The physical conclusions ($g_m^2 \propto 1/\alpha_D$) are independent of this choice.} ($n \in \mathbb{Z}$). The basis for calculating non-separable scattering amplitudes and subsystem entanglement entropy is established by this non-perturbative vertex structure.

\section{Scattering Amplitude and Reduced Density Matrix:}\label{scattering}
We monitor the evolution of the bipartite state consisting of the visible-sector fermion and the dark magnetic monopole in the joint Hilbert space $\mathcal{H}_{\text{vis}} \otimes \mathcal{H}_{\text{dark}}$ \cite{Taylor:1972pty,Kowalska:2024kbs} in order to rigorously assess the quantum correlations produced during the scattering process.

\subsection{Unitary Evolution and S-Matrix Formalism}
Let us use a direct product state as the initial asymptotic state of the composite system prior to the scattering event:
\bea
\vert{}\Psi_i\rangle = \vert{}\psi_{\text{vis}}(p_1, s_1)\rangle \otimes \vert{}\Phi_{\text{dark}}(p_2)\rangle.
\eea
Under the exact unitary scattering operator $S = \mathbb{I} + i T$, the initial product state becomes a non-separable entangled final state \cite{Weinberg:1995mt}, where $I$ is the identity operator describing free, unscattered propagation and $T$ is the transition operator encoding the non-trivial interaction dynamics mediated by the dyonic vertex and the kinetic-mixing portal. The formal superposition is obtained by applying the $S$-operator on the initial product state $\vert{}\Psi_i\rangle$. 
\bea
\vert{}\Psi_f\rangle = S\vert{}\Psi_i\rangle = (I + iT) \vert{}\Psi_i\rangle = \vert{}\Psi_i\rangle + i T \vert{}\Psi_i\rangle.
\eea 
 We introduce a whole set of multi-particle intermediate states spanning the joint Hilbert space to express this final state on a physically observable basis. This necessitates integrating across the relativistic invariant phase space for the outgoing visible fermion, parametrized by the Lorentz-invariant measure $\frac{d^3p_3}{(2\pi)^3 2E_3}$, and summing over all permitted discrete final-state channels $n$ and spin configurations $s_3$. The probability amplitudes for moving from the initial state into a particular final configuration comprised of an outgoing visible fermion state $\vert{}\psi_{\text{vis}}^{(n)}(p_3, s_3)\rangle$ entangled with a corresponding dark-sector state $\vert{}\phi_{\text{dark}}^{(n)}\rangle$ are represented by the resulting projection coefficients, $C_n(p_3, s_3)$. These coefficients provide the mathematical basis for building the bipartite density matrix and assessing subsequent quantum correlations since they are directly proportional to the invariant transition matrix element $\mathcal{M}_{fi}$ defined by the Feynman rules of the extended dark-photon and magnetic-monopole framework. The whole entangled final state is obtained by substituting this dynamical amplitude back into the formal operator expansion as follows:
\bea
\vert{}\Psi_f\rangle = S\vert{}\Psi_i\rangle = \sum_{n,s_3} \int \frac{d^3p_3}{(2\pi)^3 2E_3} C_n(p_3, s_3) \vert{}\psi_{\text{vis}}^{(n)}(p_3, s_3)\rangle \otimes \vert{}\phi_{\text{dark}}^{(n)}\rangle.
\eea
This direct proportionality provides the precise mathematical basis required to assess subsystem entanglement and decoherence rates by establishing the vital link between Feynman-rule computations of $\mathcal{M}_{fi}$ and the structural building of the bipartite density matrix. The transition matrix element mediated by the massive dark photon exchange for a visible fermion (momentum $p_1$) scattering off a dark magnetic monopole (mass $M$, magnetic charge $g_m = \frac{2\pi n}{g_D}$) is given by \cite{Kazama:1976fm,Milton:2006cp}:
\bea
\mathcal{M}_{fi} = -\epsilon g_1 Y_f g_m \, \bar{u}(p_3, s_3) \gamma^\mu u(p_1, s_1) \, \frac{-i \left( g_{\mu\nu} - \frac{q_\mu q_\nu}{m_{A'}^2} \right)}{t - m_{A'}^2} \, J_{\text{mono}}^\nu(q).
\eea
The momentum transfer squared $t = (p_1 - p_3)^2$ that appears in the propagator denominator must be explicitly related to the scattering geometry in order to assess the angular dependency and momentum scaling of this exchange. The initial-state momentum $p_{\text{cm}}$ and the scattering angle $\theta_{\text{cm}}$ explicitly determine this invariant in the center-of-mass frame through the relation;
\bea
t = -2 \, p_{\text{cm}}^2 \left( 1 - \cos\theta_{\text{cm}} \right).
\eea
The momentum-transfer suppression of the enormous dark photon propagator $(t - m_{A'}^2)^{-1}$ is explicitly incorporated into the transition amplitudes in this kinematic mapping. The denominator dictates the energy
dependence of the cross-section across distinct kinematic regime. The dark photon mass scale dominates the propagator in the soft or forward-scattering regime where $\vert{}t\vert{} \ll m_{A'}^2$, saturating at $-1/m_{A'}^2$ to prevent extreme infrared Coulomb divergences. On the other hand, the amplitude changes into a $1/\vert{}t\vert{}$ power-law suppression in the hard-scattering regime when massive momentum transfers fulfill $\vert{}t\vert{} \gg m_{A'}^2$. As a result, the monopole current's explicit form is parameterized as follows:
\bea
J^\nu_{\text{mono}}(q) = g_m F(q^2) v^\nu(q),
\eea
where $v^\nu(q)$ signifies the kinematic factor corresponding to the monopole four-velocity, which approaches $v^\nu \approx (1, 0, 0, 0)$ in the non-relativistic limit, and $g_m = \frac{4\pi n}{g_D}$ represents the topological magnetic charge. 
Together with the non-perturbative core form factor $F(q^2)$ embedded in the monopole current $J_{\text{mono}}^\nu(q)$, this strong propagator structure guaranties that the ensuing entanglement entropy evaluations and differential cross-sections remain ultraviolet finite at all energy scales. Additionally, the non-perturbative core form factor of the soliton must be encoded in the monopole current $J^\nu_{\text{mono}}(q)$ in order to account for the extended topological structure of the magnetic monopole and guaranty a well-behaved high-energy limit.  A topological magnetic monopole, in contrast to elementary point particles, has a finite core size $r_{\text{core}} \sim 1/M_W$, which causes a momentum-dependent suppression for massive momentum transfers. The core form factor $F(q^2)$ is defined via the spatial Fourier transform of the internal monopole core density $\rho_{\text{core}}(x)$, expressed as;
\bea
F(q^2) = \int d^3x \, e^{i q \cdot x} \rho_{\text{core}}(x),
\eea
satisfying the normalization condition $F(0) = 1$ to cleanly recover the standard point-particle limit for small momentum transfers where $\vert{}q\vert{} \ll M_W$. The form factor provides the required ultraviolet suppression for large momentum transfers where $\vert{}q^2\vert{} \gg M_W^2$. Typical theoretical treatments use either a dipole form $F(q^2) = (1 + q^2 / M_W^2)^{-2}$ or a Gaussian parametrization $F(q^2) = \exp(-q^2 / M_W^2)$. By explicitly incorporating the non-perturbative core structure, the total scattering cross-section is guaranteed to remain finite at high energies, directly translating the abstract form factor discussion into a concrete, calculable tool for phenomenology.

\subsection{Partial Trace and Construction of the Reduced Density Matrix}\label{Partial trace}
Since physical detectors can only access the visible Standard Model (SM) subspace $\mathcal{H}_{\text{vis}}$, the unobserved dark monopole and hidden radiation states belonging to $\mathcal{H}_{\text{dark}}$ are traced out \cite{Nielsen:2012yss}. The quantum state of the visible subsystem is therefore described by the reduced density matrix $\rho_{\text{vis}}$:
\bea
\rho_{\text{vis}} = \operatorname{Tr}_{\text{dark}} \left( \vert{}\Psi_f\rangle \langle \Psi_f\vert{} \right).
\eea

The diagonal approximation is used to calculate $\rho_{\text{vis}} = \operatorname{Tr}_{\text{dark}} \left( \vert{}\Psi_f\rangle \langle \Psi_f\vert{} \right)$. The off-diagonal momentum components disappear when $p_3 \neq p_3'$. Two complementary frameworks can be used to comprehend this. From the decoherence-based perspective, after the scattering event, the visible fermion inevitably interacts with the surrounding environment and detector media, causing rapid environmental decoherence between different momentum states. Thus, only the diagonal components where $p_3 = p_3'$ remain after the off-diagonal elements $\langle p_3 \vert{} \rho_{\text{vis}} \vert{} p_3' \rangle$ decay exponentially on a characteristic timeframe $\tau_{\text{decoherence}} \ll t_{\text{obs}}$. On the other hand, from a measurement-based point of view, a genuine scattering experiment consists of a detector that records the definite final momentum $p_3$ of the visible fermion. By projecting the density matrix onto momentum eigenstates, this measurement technique guaranties that the post-measurement density matrix is diagonal in the momentum representation and efficiently traces over off-diagonal coherences in the momentum basis. Additionally, when evaluating unpolarized cross-sections, spin coherences ($s_3 \neq s_3'$) are explicitly averaged over (or kept for polarization-resolved studies), completing the precise structure of $\rho_{\text{vis}}$.

The partial trace operation yields the following when the expanded final state is substituted:
\bea
\rho_{\text{vis}} = \int \frac{d^3 p_3}{(2\pi)^3 2E_3} \sum_{n} \left\vert{} \mathcal{C}_n(p_3) \right\vert{}^2 \, \vert{}\psi_{\text{vis}}^{(n)}(p_3)\rangle \langle \psi_{\text{vis}}^{(n)}(p_3) \vert{}.
\eea

By executing a partial trace over the unobserved dark-sector states from the pure bipartite final density matrix, this equation defines the reduced density matrix $\rho_{\text{vis}}$ of the visible subsystem. The Lorentz-invariant phase space measure is given by the integral $\int \frac{d^3p_3}{(2\pi)^3 2E_3}$, which sums together all kinematically permitted final-state momenta $p_3$ for the outgoing visible fermion. Contributions from any internal quantum channels, spin configurations, or intermediate states indexed by $n$ are combined inside this integral via the discrete sum $\sum_n$. The transition matrix elements $\mathcal{M}_{fi}$ are directly linked to the squared absolute values of the expansion coefficients, $\vert{}C_n(p_3)\vert{}^2$, which serve as the physical probability weights controlling each channel. Lastly, the pure-state density operators for the visible fermion are constructed by the outer product projectors $\vert{}\psi_{\text{vis}}^{(n)}(p_3)\rangle\langle\psi_{\text{vis}}^{(n)}(p_3)\vert{}$, which show how tracing out the unmeasured dark sector converts the pure global state into a statistical mixture of visible states weighted by their respective scattering probabilities.\\
The S-matrix expansion $S = 1 + iT$ must be substituted into the bipartite final-state definition in order to go from the general momentum-space expression to the perturbative form. Any quadratic term in $C_n(p_3)$ naturally scales as $\epsilon^2$ because the state expansion coefficients $C_n(p_3)$ are directly proportional to the transition matrix element $\mathcal{M}_{fi}$, which scales linearly with the kinetic-mixing parameter $\epsilon$. It is possible to explicitly reparameterize the Lorentz-invariant phase-space integral $\int \frac{d^3p_3}{(2\pi)^3 2E_3}$ in the center-of-mass frame into an angular integration over the solid angle $d\Omega_{\text{cm}}$ and an energy delta-function component. The projector $\vert{}\psi_i\rangle\langle\psi_i\vert{}$ is obtained when the identity operator piece of $S = 1$ acts on the initial state, isolating the unscattered portion. This unscattered term is attenuated by $(1 - \Gamma_{\text{int}})$, where $\Gamma_{\text{int}}$ represents the total integrated interaction rate proportional to $\epsilon^2$, in order to meet total probability conservation under unitary time evolution up to $\mathcal{O}(\epsilon^2)$. On the other hand, the real scattering contributions are produced by the dynamic $iT$ components. Summing over all intermediate dark channels $n$ and substituting the squared matrix elements containing the massive dark photon propagator and kinematic factors gathers everything into the kernel $K(s, t, m_{A'}, M)$. This kernel then multiplies the scattered state projector $\vert{}\psi_{\text{scat}}\rangle\langle\psi_{\text{scat}}\vert{}$ under the remaining solid-angle integral, while pushing higher-order multi-exchange contributions into the $\mathcal{O}(\epsilon^4)$ remainder.
Extending the kinetic-mixing parameter $\epsilon^2$ to the leading non-trivial order:
\bea
\rho_{\text{vis}} = \left( 1 - \Gamma_{\text{int}} \right) \vert{}\psi_i\rangle \langle \psi_i \vert{} + \epsilon^2 \int d\Omega_{\text{cm}} \, \mathcal{K}\left(s, t, m_{A'}, M\right) \vert{}\psi_{\text{scat}}\rangle \langle \psi_{\text{scat}} \vert{} + \mathcal{O}(\epsilon^4),
\eea
where $\mathcal{K}$ contains the kinematic phase-space weighting controlled by the monopole partial-wave expansion, and $\Gamma_{\text{int}} = \int \frac{d\sigma_{\text{dyon}}}{d\Omega} d\Omega$ represents the entire integrated dyonic interaction probability. The function $K(s, t, m_{A'}, M)$ is explicitly described in terms of the unpolarized differential cross-section for dyonic monopole-fermion scattering in order to guarantee total reproducibility and clarify the phase-space weighting governed by the monopole expansion. When the kinetic-mixing portal coupling $\epsilon^2$ is factored out, the kinematic weighting function is as follows:
\bea
K(s, t, m_{A'}, M) \equiv \frac{1}{\epsilon^2} \frac{p_{\text{cm,initial}}}{4\pi} \frac{d\sigma_{\text{dyon}}}{d\Omega_{\text{cm}}},
\eea
where $p_{\text{cm,initial}}$ denotes the center-of-mass momentum of the initial state, and the differential cross-section is expressed as follows:
\bea
\frac{d\sigma_{\text{dyon}}}{d\Omega_{\text{cm}}} = \frac{1}{64\pi^2 s} \frac{\epsilon^2 g_1^2 Y_f^2 g_m^2}{(t - m_{A'}^2)^2} L_{\mu\nu} H^{\mu\nu}.
\eea
Here, $t = (p_1 - p_3)^2 = -2 p_{\text{cm}}^2 (1 - \cos\theta_{\text{cm}})$ represents the momentum transfer squared in the center-of-mass frame. The spin-averaged lepton tensor $L_{\mu\nu}$ and the monopole tensor $H^{\mu\nu}$ are defined, respectively, as follows;
\bea
L_{\mu\nu} = \frac{1}{2} \sum_{s_1, s_3} \left[ \bar{u}(p_3, s_3)\gamma_\mu u(p_1, s_1) \right] \left[ \bar{u}(p_1, s_1)\gamma_\nu u(p_3, s_3) \right],
\eea
and
\bea
H^{\mu\nu} = J^\mu_{\text{mono}}(q) J^{*\nu}_{\text{mono}}(q),
\eea
where the bar denotes standard spin summing and averaging over initial and final degrees of freedom, thereby enabling direct verification of the parametric scaling and numerical evaluations.\\

Once the explicit form of the reduced density matrix has been established, the framework easily moves on to assessing the quantum entanglement created between the visible fermion and the unobserved dark sector. The statistical uncertainty that results from tracing out the dark-sector degrees of freedom is a direct reflection of quantum information loss into the hidden sector since it converts a pure global state into the mixed density matrix $\rho_{\text{vis}}$. The off-diagonal coherences and purity loss of $\rho_{\text{vis}}$ directly measure the backreaction and quantum entanglement generated by the hidden-sector portal.

\section{Quantum Entanglement Metrics:} \label{entropy}
We investigate the information-theoretic properties of the reduced density matrix $\rho_{\text{vis}}$ \cite{Nielsen:2012yss,Horodecki:2009zz} to accurately measure the quantum links between the isolated dark monopole and the visible fermion throughout the scattering process.

\subsection{Von Neumann Entanglement Entropy}
The primary diagnostic method for identifying if a bipartite quantum state is non-separable is the von Neumann entropy of the reduced density matrix \cite{Plenio:2007zz} which is calculated as follows:
\bea 
S_{\text{ent}} = -\operatorname{Tr}_{\text{vis}} \left( \rho_{\text{vis}} \log_2 \rho_{\text{vis}} \right).
\eea \label{sent}
The degree of quantum entanglement and mixedness produced by tracing out unobserved dark-sector degrees of freedom is quantified by Equation~\ref{sent}, which also specifies the von Neumann entropy for the visible subsystem. Also defined as entanglement entropy, which quantifies the amount of quantum information leaked from the visible fermion into the hidden sector, is represented by the $S_{\text{ent}}$ and is expressed in bits (because of the base-2 logarithm). $S_{\text{ent}} = 0$ for an initial state that is entirely pure. The probability distribution that results from transitions into the dark sector caused by kinetic-mixing interactions ($\epsilon$) broadens across several channels, increasing $S_{\text{ent}}$ and offering a precise measure of environmental decoherence and quantum information loss. The trace of the mixed density matrix simplifies to a sum over its diagonal eigenvalues $\lambda_n$ when it is written in its orthonormal eigenbasis. The spectrum splits into the dynamic scattering channels and the unscattered baseline. The unscattered mode, $\lambda_0$, represents the likelihood that the visible fermion will pass through the target without interacting. It is defined as $\lambda_0 = 1 - \Gamma_{\text{int}}$, where the total integrated dyonic interaction probability is represented as $\Gamma_{\text{int}} = \int \frac{d\sigma_{\text{dyon}}}{d\Omega} d\Omega$. Within the hidden sector Hilbert space $\mathcal{H}_{\text{dark}}$, $\lambda_k$ represents the scattered channels that indicate the likelihood of transitioning into particular orthogonal states. The partial-wave distribution $\chi_k$ governs these, which are parameterized as $\lambda_k = \epsilon^2 \chi_k(s, m_{A'}, M)$, scaling quadratically with the kinetic-mixing parameter $\epsilon$.
The completeness relation requires $\lambda_0 + \sum_{k=1}^\infty \lambda_k = 1$, which implies: $\sum_{k=1}^\infty \epsilon^2 \chi_k = \Gamma_{\text{int}}$.
\bea
\lambda_0 = 1 - \Gamma_{\text{int}}, \quad \lambda_k = \epsilon^2 \chi_k(s, m_{A'}, M),
\eea
where $\chi_k$ is the partial-wave probability distribution over the hidden sector Hilbert space $\mathcal{H}_{\text{dark}}$ and $\Gamma_{\text{int}} = \int \frac{d\sigma_{\text{dyon}}}{d\Omega} d\Omega$ is the total integrated dyonic interaction probability. Substituting these explicit eigenvalues into the trace definition $S_{\text{ent}} = -\text{Tr}_{\text{vis}} (\rho_{\text{vis}} \log_2 \rho_{\text{vis}})$ yields the expanded discrete formula:
\bea
S_{\text{ent}} = -\lambda_0 \log_2 \lambda_0 - \sum_{k=1}^{\infty} \lambda_k \log_2 \lambda_k.
\eea
The logarithm base conversion identity $\log_2 x = \frac{\ln x}{\ln 2}$ is applied in order to scale the entire statement by $1/\ln 2$ in order to move from the discrete entropy summation to the final parametric scaling relation. The explicit eigenvalue definitions are then entered into the equation, with $\lambda_0 = 1 - \Gamma_{\text{int}}$ for the unscattered mode and $\lambda_k = \epsilon^2 \chi_k$ for the scattered channels. The Taylor series approximation $\ln(1 - x) \approx -x$ is used for the unscattered baseline term when the interaction probability ($\Gamma_{\text{int}} \ll 1$) are minimal. This decreases the unscattered contribution $-(1 - \Gamma_{\text{int}}) \ln(1 - \Gamma_{\text{int}})$ to roughly $(1 - \Gamma_{\text{int}}) \Gamma_{\text{int}}$, which reduces to $\Gamma_{\text{int}}$ at leading order. In parallel, the logarithmic product is broken into $\ln(\epsilon^2) + \ln(\chi_k)$ to expand the infinite summation over the distributed channels, $-\sum_{k=1}^{\infty} \epsilon^2 \chi_k \ln(\epsilon^2 \chi_k)$. This summation splits into a dominant term proportional to $-\Gamma_{\text{int}} \ln(\epsilon^2)$ alongside the internal channel distribution sum because the normalization condition requires that the sum of the partial-wave probability distribution equals the total integrated interaction probability ($\sum \epsilon^2 \chi_k = \Gamma_{\text{int}}$). The overall interaction probability is factorized out of the bracket by combining these expanded portions, resulting in $\frac{\Gamma_{\text{int}}}{\ln 2} \left[ \ln e - \ln(\epsilon^2) \right] - \frac{1}{\ln 2} \sum_{k=1}^\infty \epsilon^2 \chi_k \ln(\chi_k)$. The internal expression is condensed by rewriting the constant $1$ as $\ln e$. Standard logarithm subtraction rules then combine $\ln e - \ln(\epsilon^2)$ into a single fraction inside the logarithm: $S_{\text{ent}} \simeq \frac{\Gamma_{\text{int}}}{\ln 2} \ln\left(\frac{e}{\epsilon^2}\right) - \frac{1}{\ln 2} \sum_{k=1}^\infty \epsilon^2 \chi_k \ln(\chi_k)$. In the limit of small kinetic mixing ($\epsilon \ll 1$), the first term scales as $\mathcal{O}(\epsilon^2 \ln(1/\epsilon^2))$. Because $\ln(1/\epsilon^2) \gg 1$ as $\epsilon \to 0$, the log-enhanced term dominates completely, rendering the second term a subleading $\mathcal{O}(\epsilon^2)$ correction that can be dropped in leading-order analytic approximations. The entanglement entropy produced during a dyonic scattering process via a kinetic-mixing portal between a visible fermion and an unobserved dark sector can be approximated analytically as follows:
\bea\label{sent}
S_{\text{ent}} \simeq \frac{1}{\ln 2} \left[ \int \frac{d\sigma_{\text{dyon}}}{d\Omega} d\Omega \right] \ln \left( \frac{e}{\epsilon^2 } \right).
\eea
Equation~\ref{sent} describes the way phase-space integration and contact strength affect how quantum information escapes the visible subsystem. Entanglement generation decreases for very weak portal couplings, as seen by the prefactor $\epsilon^2$, which represents the quadratic suppression of the kinetic-mixing portal. The basic conversion term to express the resulting entropy in standard bits of information is $1/\ln 2$. Additionally, the overall scattering cross-section across all solid angles is explained by the total integrated dyonic interaction probability, which is given by the integral $\int \frac{d\sigma_{\text{dyon}}}{d\Omega} d\Omega$. Lastly, the internal logarithmic expression captures the relative weight of available phase-space channels in relation to the unscattered baseline by acting as a structural enhancement factor that grows inversely with the total interaction rate.

\subsection{Parametric Scaling and Topological Amplification}
The entanglement entropy exhibits a typical logarithmic enhancement $\epsilon^2 \ln(1/\epsilon^2)$ \cite{Kowalska:2024kbs}, whereas ordinary perturbative cross-sections scale simply as $\mathcal{O}(\epsilon^2)$. Two different physical behaviors controlled by the kinetic-mixing portal and the dark sector topology are shown by the analytical structure of $S_{\text{ent}}$: $\epsilon^2 \ln(1/\epsilon^2)$ Scaling. This logarithmic adjustment originates from the multi-channel trace across the unrecorded hidden-sector radiation modes. Topological Monopole Enhancement: Using the previously computed explicit total dyonic cross-section $\sigma_{\text{tot}}(s)$ \cite{Preskill:1984gd,Goddard:1977da}:
\bea
S_{\text{ent}} \simeq \frac{1}{\ln 2} \frac{\pi \epsilon^2 \alpha Y_f^2 n^2}{s \alpha_D} \left[ \frac{2s^2 + 3s m_{A'}^2 + 2m_{A'}^4}{m_{A'}^2(s + m_{A'}^2)} - 2\left(1 + \frac{m_{A'}^2}{s}\right)\ln\left(1 + \frac{s}{m_{A'}^2}\right) \right] \ln \left( \frac{e}{\epsilon^2 } \right).
\eea
The magnetic charge $g_m = \frac{n}{2 e_D}$ implies $g_m^2 \propto \frac{n^2}{\alpha_D}$, indicating that a smaller dark fine-structure constant $\alpha_D$ (weak electric coupling) or large topological winding numbers $n$ significantly enhance magnetic charge and associated quantum entanglement generation.

\subsection{Purity and Linear Entropy Bounds}
In terms of the purity of the reduced density matrix $\gamma = \operatorname{Tr}_{\text{vis}}(\rho_{\text{vis}}^2)$, we compute the linear entropy $S_{\text{lin}}$ \cite{Zyczkowski:2006abi}, to supplement the von Neumann entropy. By using the trace of the squared density matrix, which is formally defined as follows, instead of the matrix logarithm, linear entropy ($S_{\text{lin}}$) offers an alternative, computationally efficient measure of quantum purity and mixedness, defined mathematically as follows:
\bea
S_{\text{lin}} = 1 - \operatorname{Tr}_{\text{vis}}(\rho_{\text{vis}}^2) = 1 - \sum_k \lambda_k^2.
\eea
Substituting the perturbed eigenvalue spectrum:
\bea
S_{\text{lin}} = 1 - \left( (1 - \Gamma_{\text{int}})^2 + \sum_{k=1}^{\infty} \left( \frac{\epsilon^2}{\ln 2} \chi_k \right)^2 \right) \approx 2 \, \Gamma_{\text{int}} + \mathcal{O}(\epsilon^4).
\eea
Since $\Gamma_{\text{int}} \propto \epsilon^2 $, the linear entropy provides a direct algebraic measure of purity loss that scales linearly with the overall dyonic scattering probability. This demonstrates that the portal coupling strength, $\epsilon^2$ directly correlates with the growth in subsystem decoherence and entanglement.

\section{Phenomenological Implications:}
When information-theoretic entanglement measures are converted into observable particle physics signatures, there are obvious consequences for high-precision scattering experiments, collider missing-energy searches, and cosmological or astrophysical probes of secluded sectors \cite{Essig:2013lka,Alexander:2016aln}.

\subsection{Decoherence and Effective Damping in Visible Scattering}
When experimental detectors measure visible-sector fermions without resolving the recoiling dark magnetic monopole or associated hidden-sector radiation, the composite system exhibits open quantum behavior \cite{Zurek:2003zz,Giulini:1996nw}. As the visible fermion moves from a pure to a mixed state, the decreased density matrix $\rho_{\text{vis}}$, which characterizes its viewed state, displays an effective decoherence phenomena. The off-diagonal coherence components of $\rho_{\text{vis}}$ degrade exponentially in time or spatial propagation, depending on the decoherence factor:
\bea
\langle \psi_{\text{vis}}^{(1)} \vert{} \rho_{\text{vis}} \vert{} \psi_{\text{vis}}^{(2)} \rangle(t) = \langle \psi_{\text{vis}}^{(1)} \vert{} \rho_{\text{vis}} \vert{} \psi_{\text{vis}}^{(2)} \rangle(0) \, \exp\left( -\Gamma_{\text{dec}} \, t_{\text{obs}} \right),
\eea
where the rate of entanglement production with the unobserved hidden sector drives the macroscopic decoherence rate $\Gamma_{\text{dec}}$:
\bea
\Gamma_{\text{dec}} \sim n_{\text{target}} \, v_{\text{rel}} \, \sigma_{\text{tot}}(s).
\eea
This formula explains how environmental decoherence causes exponential suppression of quantum coherence inside the visible subsystem during time or spatial propagation. The quantum superposition between different visible states $\vert{}\psi_{\text{vis}}^{(1)}\rangle$ and $\vert{}\psi_{\text{vis}}^{(2)}\rangle$ is represented by the off-diagonal elements of the reduced density matrix, represented by $\langle\psi_{\text{vis}}^{(1)} \vert{} \rho_{\text{vis}} \vert{} \psi_{\text{vis}}^{(2)} \rangle$. These off-diagonal terms control phase relationships and interference phenomena, in contrast to diagonal elements that follow conventional probability weights. Unobserved dark-sector degrees of freedom function as an external environment that constantly observes and entangles with the visible system when the visible fermion interacts with the hidden sector through the kinetic-mixing portal. The rate at which this phase information leaks into the unobserved environment is determined by the decoherence rate $\Gamma_{\text{dec}}$, and $t_{\text{obs}}$ is either the elapsed observation time or the spatial propagation distance scaled by velocity. The quantum superposition quickly collapses into a classical statistical mixture as time increases due to the exponential damping factor $\exp(-\Gamma_{\text{dec}} t_{\text{obs}})$ driving the off-diagonal terms toward zero. As a direct result of continuous particle leakage and hidden-sector interactions, this environmental decoherence mechanism explains why quantum interference effects vanish on macroscopic scales, turning pure quantum states into essentially classical ensembles.
Because $\sigma_{\text{tot}}(s) \propto \epsilon^2$, this phase damping provides minor alterations to forward-scattering phase shifts, polarization asymmetries, or quantum interference patterns in precise high-energy collisions. As a result, an indirect observable signal is produced that is not dependent on conventional resonance searches.

\subsection{Infrared Entanglement Signatures from Soft Dark Radiation}
Particles emit soft dark photon ($A^\prime$) bremsstrahlung during the intense acceleration phase of a monopole-fermion collision, which is below the experimental energy resolution threshold $\Delta E$. These particles escape unmeasured into an unobserved dark sector and drive an infrared-divergent entanglement shift across the bipartite boundary \cite{Bloch:1937pw,Weinberg:1965nx}.  The quantum soft theorem factorizes the emission amplitude into a universal classical eikonal current, controlled by the fermion's incoming ($p_1$) and outgoing ($p_3$) momenta, multiplied by the hard core dyonic scattering matrix element ($\mathcal{M}_{\text{hard}}$), in order to quantify this open quantum system process without having to solve intractable multi-particle diagrams. By explicitly isolating the severe singularities that arise when the radiated momentum disappears ($k \to 0$), this mathematical separation offers the analytical handle needed to couple virtual corrections with real soft emissions and provide finite, physically consistent entanglement entropy estimates.
Applying the quantum soft theorem to open bipartite systems \cite{Strominger:2017zoo}, the emission amplitude factorizes into the soft current operator acting on the hard dyonic scattering amplitude.
\bea
\mathcal{M}_{\text{soft}+\text{hard}} \approx g_D q_f \left( \frac{p_3 \cdot \epsilon^*}{p_3 \cdot k} - \frac{p_1 \cdot \epsilon^*}{p_1 \cdot k} \right) \mathcal{M}_{\text{hard}}.
\eea
The entanglement entropy is increased by a logarithmic infrared adjustment when integrating over the phase space of unresolved soft dark photons up to the cutoff $\Delta E$:
\bea
\Delta S_{\text{ent}}^{\text{soft}} = \frac{\epsilon^2 g_1^2 Y_f^2 g_D^2}{4\pi^2} \ln\left(\frac{E_{\text{cm}}}{\Delta E}\right) \ln\left(\frac{m_{A'}^2 + \mathbf{q}^2}{m_{A'}^2}\right).
\eea
This logarithmic correction to the entanglement entropy ($\Delta S_{\text{ent}}^{\text{soft}}$) is obtained by integrating the phase space of unresolved soft dark photon emissions up to the experimental energy resolution threshold $\Delta E$, where the coupling prefactor ($\epsilon^2 g_1^2 Y_f^2 g_D^2 / 4\pi^2$) scales the strength of the hidden-sector portal interaction. The traditional infrared divergence resulting from summing over soft modes below the detector's resolution is captured by the first logarithmic term, $\ln(E_{\text{cm}}/\Delta E)$, which scales with the center-of-mass energy in relation to the cutoff threshold. Concurrently, the dark photon mass ($m_{A^\prime}$) and momentum transfer squared ($q^2$) are included in the second logarithmic term, which functions as a physical regulator to control potential singularities. Together, these variables quantify the logarithmic growth of entropy caused by unmeasured soft radiation leaking into the dark sector, formally connecting macroscopic quantum information loss across the bipartite border to microscopic gauge parameters.
This infrared structure reveals a quantitative relationship between quantum entanglement entropy and soft-photon theorems and demonstrates that soft hidden-sector gauge dynamics are directly stored in subsystem purity loss, much like traditional Faddeev-Popov cancelation approaches.

\subsection{Complementary Probes Beyond Traditional Cross-Sections}
Beam-dump experiments, direct detection limits, and precision electroweak observables are examples of conventional constraints on kinetic-mixing portals that simply rely on inclusive decay widths and total cross-sections scaling as $\mathcal{O}(\epsilon^2)$ \cite{Holdom:1985ag,Jaeckel:2010ni}. Despite their robustness, by integrating over all final-state configurations, these inclusive measures lose quantum phase information.\\

There are several obvious advantages of measuring entanglement entropy and purity loss as an extra portal probe.
\begin{itemize}
\item The way that quantum information metrics capture profound non-perturbative physics that conventional perturbative cross-sections miss is highlighted by their remarkable sensitivity to topological configurations. When the system encounters strong couplings or non-trivial topological winding numbers, the magnetic charge $g_m$ scales inversely with the secluded-sector coupling ($\alpha_D$), and its squared contribution ($g_m^2 \propto 1/\alpha_D$) drives a quadratic amplification of the soft entanglement entropy. Monopole-fermion collisions are essentially controlled by Dirac string singularities and global topological invariants that determine the spatial geometry of the interaction, whereas traditional perturbative approaches assume trivial vacuum structures and weak interactions. Monitoring these quantum information metrics provides a sensitive probe into strong secluded-sector dynamics that are totally undetectable to conventional scattering amplitude analyzes \cite{Preskill:1984gd}, revealing how topological background fields modify soft dark photon radiation.

\item Degeneracy Breaking: This phenomenon resolves a crucial ambiguity in dark photon mass generation: in certain kinematic limits, different microphysical frameworks, such as the dynamical dark Higgs mechanism (which has physical scalar radial modes) versus the Stueckelberg mechanism (which does not), produce identical inclusive scattering cross-sections. The partial-wave entanglement spectrum effectively breaks this degeneracy, but conventional cross-section measurements are unable to differentiate between these scenarios. Physical scalar modes provide unique, resolvable traces on quantum information metrics such as entanglement entropy because they modify the internal channel structure and phase-space distribution of radiated soft quanta. As a result, quantum bipartite measurements offer a potent diagnostic tool that may examine UV-complete dark sector mass formation mechanisms that are completely obscured by conventional perturbative scattering observables \cite{Pospelov:2007mp}.

\item The allowed parameter space of the kinetic mixing parameter ($\epsilon$), dark photon mass ($m_{A'}$), and secluded-sector coupling ($\alpha_D$) are strictly limited by information-theoretic bound constraints, which demand that the entanglement entropy produced during scattering events adhere to basic physical unitarity limits. Unobserved parameter configurations may theoretically violate probability conservation principles or core quantum information limitations because unobserved soft dark photon emissions produce logarithmic shifts in entanglement entropy across the bipartite boundary. By enforcing these information-theoretic limits, theoretical models are kept physically feasible and quantum entanglement metrics become an autonomous, potent diagnostic tool that limits the phenomenology of hidden-sector portals beyond the capabilities of conventional cross-section measurements \cite{Irakleous:2021ggq}.
\end{itemize}

\begin{figure}[tb]
\begin{center}
\includegraphics[width=15.0cm]{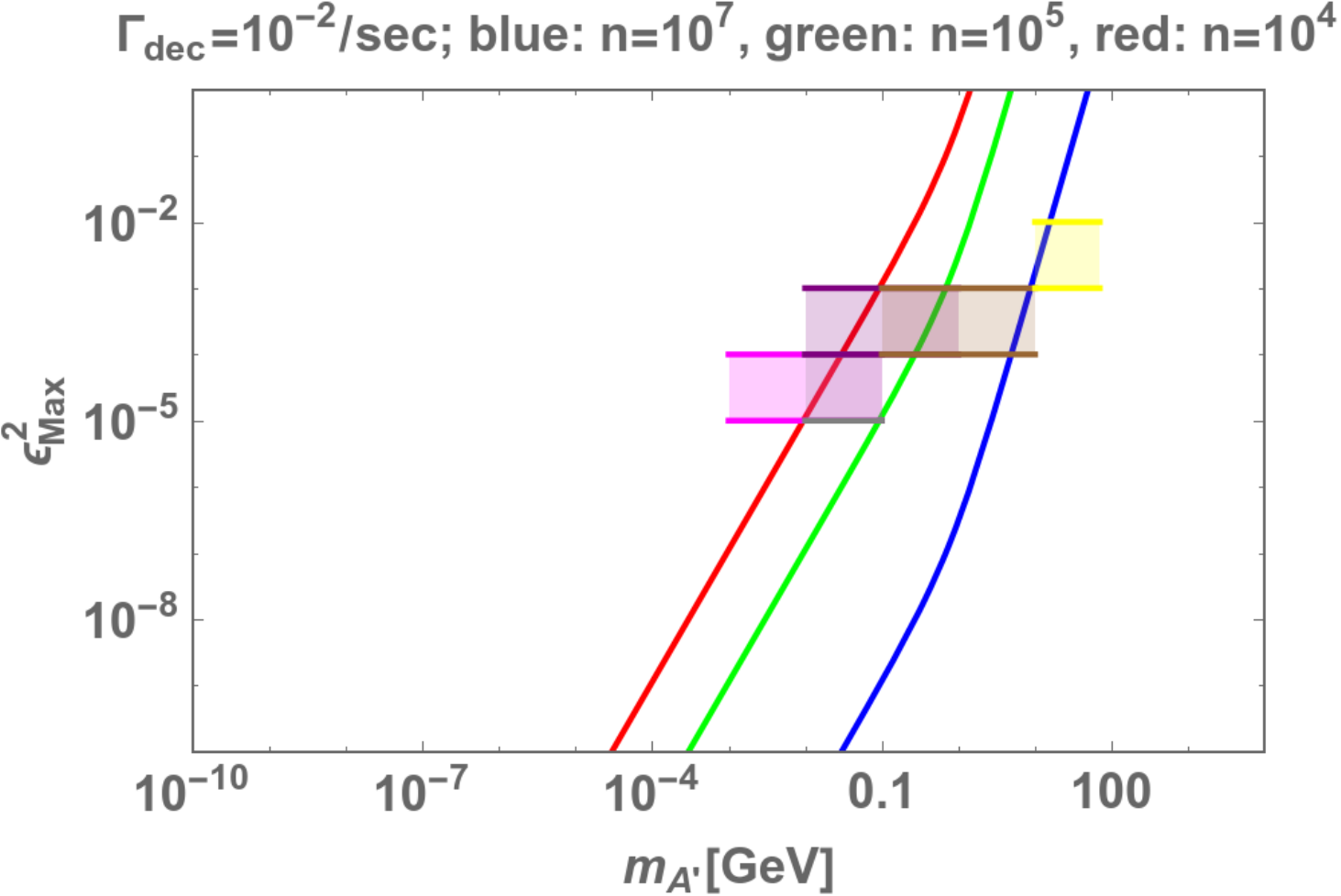}
\caption{Exclusion limits in the $(\epsilon, m_{A'})$ parameter space, where we have adopted $\Gamma_{\rm exp}=10^{-2}$ sec$^{-1}$. The colored shaded regions represent existing constraints from beam-dump and collider experiments: E137 gray, $m_{A'} \sim 0.01$--$0.1$~GeV), NA64 (magenta, $m_{A'} \sim 0.001$--$0.1$~GeV), BaBar (brown, $m_{A'} \sim 0.1$--$10$~GeV), and KLOE (purple, $m_{A'} \sim 0.01$--$1$~GeV). The curves show the maximal sensitivity of our proposed decoherence-based framework for different topological winding numbers: $n=10^7$ (blue), $n=10^5$ (green), and $n=10^4$ (red), where the ranges below these lines are allowed.
The downward shift for larger $n$ demonstrates the topological amplification effect ($\Gamma_{\text{dec}} \propto n^2$), which allows probing significantly smaller kinetic mixing parameters $\epsilon$ beyond the reach of standard perturbative cross-section measurements. 
}
  \label{fig:ma-epsilon}
\end{center}\end{figure}

Finally, we have shown exclusion limits in the $(\epsilon, m_{A'})$ parameter space in Figure.~\ref{fig:ma-epsilon}, where we have adopted the following typical experimental parameters: $\Gamma_{\rm exp}=10^{-2}$ sec$^{-1}$, $n_{\rm target}=10^{14}$ cm$^{-3}$, $v_{\rm rel}=10^5$ cm/sec, $\alpha=1/137$, $Y_f=1.0$ $\alpha_D = 0.2$, and $\sqrt s=1$ GeV. The colored shaded regions represent existing constraints from beam-dump and collider experiments: E137~\cite{Bjorken:1988as} gray, $m_{A'} \sim 0.01$--$0.1$~GeV), NA64~\cite{ArgoNeuT:2019ckq} (magenta, $m_{A'} \sim 0.001$--$0.1$~GeV), BaBar~\cite{BaBar:2014zli} (brown, $m_{A'} \sim 0.1$--$10$~GeV), and KLOE~\cite{Brauner:2016pko} (purple, $m_{A'} \sim 0.01$--$1$~GeV). The curves show the maximal sensitivity of our proposed decoherence-based framework for different topological winding numbers: $n=10^7$ (blue), $n=10^5$ (green), and $n=10^4$ (red), where the ranges below these lines are allowed.
The downward shift for larger $n$ demonstrates the topological amplification effect ($\Gamma_{\text{dec}} \propto n^2$), which allows probing significantly smaller kinetic mixing parameters $\epsilon$ beyond the reach of standard perturbative cross-section measurements

\section{Conclusion and results}
Traditional searches for hidden sectors, relying predominantly on missing-energy ($E_{\text{miss}}$) and missing-momentum signatures at high-energy colliders or beam-dump experiments, suffer from fundamental blind spots. These conventional methods are inherently inclusive, averaging over unobserved final states while completely discarding the phase-coherent quantum information and environmental backaction occurring between the visible and hidden sectors. Furthermore, traditional missing-energy paradigms are bottlenecked by high energy thresholds, background limitations, and an inability to capture sub-leading quantum correlations or cumulative purity degradation that do not manifest as hard, single-event missing energy spikes. This work overcomes these limitations by shifting the investigative paradigm from inclusive particle counting to open quantum systems and quantum information diagnostics. Rather than waiting for a rare missing-energy event at a high-energy collider, our framework demonstrates that hidden-sector interactions leave an indelible, cumulative footprint on the visible subsystem through phase damping, purity loss, and macroscopic decoherence. By treating the Standard Model visible sector as an open quantum system, we captured non-perturbative multi-channel traces that yield a distinctive logarithmic entanglement amplification scaling as $S_{\text{ent}} \simeq \frac{1}{\ln 2} [\sigma_{\text{tot}}(s)] \ln \left( \frac{e}{\epsilon^2 } \right)$, a parametric behavior entirely invisible to standard perturbative cross-section measurements. The analytical results derived here bridge microscopic field theory with macroscopic measurement, transforming abstract information-theoretic quantities into concrete experimental observables. By evaluating the explicit $t$-channel dark-photon exchange across monopole-fermion scattering, we established the dimensionally balanced total dyonic cross-section $\sigma_{\text{tot}}(s) = \frac{\pi \epsilon^2 \alpha Y_f^2 n^2}{s \alpha_D} \left[ \frac{2s^2 + 3s m_{A'}^2 + 2m_{A'}^4}{m_{A'}^2(s + m_{A'}^2)} - 2\left(1 + \frac{m_{A'}^2}{s}\right)\ln\left(1 + \frac{s}{m_{A'}^2}\right) \right]$. We proved that this microscopic scattering drives the exponential decay of off-diagonal density matrix coherence elements at a concrete rate, $\Gamma_{\text{dec}} \sim n_{\text{target}} v_{\text{rel}} \sigma_{\text{tot}}(s)$, while quantum soft theorems account for infrared dynamics via unresolved soft dark-photon emission below the detector threshold $\Delta E$, yielding the explicit logarithmic entanglement shift $\Delta S_{\text{soft}}^{\text{ent}}$. Crucially, these results open an immediate, actionable pathway for precision laboratory experiments that bypass collider energy bounds. In ultra-precise atom or molecular interferometry and high-coherence macroscopic systems passing through dense matter targets ($n_{\text{target}}$), any unobserved anomalous fringe visibility loss or coherence degradation sets a strict upper bound on the environmental decoherence rate ($\Gamma_{\text{dec}} \le \Gamma_{\text{exp}}$). This allows us to place direct quantitative constraints on the fundamental portal parameter space, yielding an exclusion limit on the direct bound of coupling to mass ratio: $\frac{\epsilon^2}{m_{A'}^2} \lesssim \frac{2 \, \Gamma_{\text{exp}}}{n_{\text{target}} \, v_{\text{rel}} \, \alpha \, Y_f^2 \, g_m^2}$. By establishing entanglement entropy, linear purity loss, and decoherence rates as rigorous experimental observables, this work provides a powerful complementary frontier to discover elusive hidden sectors through subsystem information leakage.

\section{Acknowledgement}
 HO is supported by Zhongyuan Talent (Talent Recruitment Series) Foreign Experts Project. 


\bibliography{References}
\bibliographystyle{Ref}
\end{document}